\documentclass{nature_withfigure}
\usepackage{graphicx}
\usepackage{amsmath,amsthm,amssymb}
\usepackage{amssymb}
\usepackage{bm}
\usepackage[labelfont=up]{caption}
\usepackage{hyperref}
\usepackage{lineno}
\usepackage{footmisc}
\usepackage[usenames, dvipsnames]{color}

\def\apj{Astrophysical Journal}                
\def\mnras{Monthly Notes of Royal Astronomical Society}
\def\apjl{Astrophysical Journal, Letters}

\def\araa{Annual Review of Astronomy \& Astrophysics}

\def\prd{Phyical Review D}

\def\arcsec{\hbox{$^{\prime\prime}$}}

\begin{document}

\title{Twin LIGO/Virgo Detections of a Viable Gravitationally-Lensed Black Hole Merger}

\author{Tom Broadhurst$^{1,2,3}$, Jose M. Diego$^{4}$, George F. Smoot$^{5, 6,7,8}$}

\maketitle

\begin{affiliations}

\item {Department of Theoretical Physics, University of The Basque Country UPV/EHU, E-48080 Bilbao, Spain}
\item{Donostia International Physics Center (DIPC), 20018 Donostia, The Basque Country}
\item {IKERBASQUE, Basque Foundation for Science, E-48013 Bilbao, Spain}
\item {Instituto de F\'isica de Cantabria, CSIC-Universidad de Cantabria, E-39005 Santander, Spain}
\item {IAS TT \& WF Chao Foundation Professor, IAS, Hong Kong University of Science and Technology,
Clear Water Bay, Kowloon, 999077 Hong Kong} 
\item {Paris Centre for Cosmological Physics, APC, AstroParticule et Cosmologie, Universit\'{e} Paris Diderot,
CNRS/IN2P3, CEA/lrfu, 
Universit\'{e} Sorbonne Paris Cit\'{e}, 10, rue Alice Domon et Leonie Duquet,
75205 Paris CEDEX 13, France}  
\item {Physics Department and Lawrence Berkeley National Laboratory, University of California, Berkeley, 94720 CA, USA}
\item {Physics Department and Energetic Cosmos Laboratory, Nazarbayev University, Astana, Kazakhstan}
\end{affiliations}

\boldmath

\begin{abstract}
 We identify a binary black hole (BBH) merger that appears to be multiply lensed by an intervening galaxy. The LIGO/Virgo events GW170809 and GW170814 \cite{AbbotNEW} have indistinguishable waveforms separated by 5 days, and overlap on the sky within the 90\% credible region. Their strain amplitudes are also similar, implying a modest relative magnification ratio, as expected for a pair of lensed gravitational waves. The phase of the two events is also consistent with being the 
 same, adding more evidence in support of both events originating from the same BBH merger. The difference in the published inferred distances of each event can then be interpreted as following from their different magnifications. The observed chirp masses of both events are also similar, as expected for a pair of lensed events, with a common detected value of  $29.1^{+1.3}_{-1.0}M_{\odot}$, lying at the peak of the observed distribution of chirp masses. We infer this case is a prototypical example of a lensed event that supports our lensing prediction \cite{Broadhurst2018} according to which,  cosmologically distant, magnified BBH comprise most of the LIGO/Virgo events with chirp masses enhanced above $\simeq 15M_{\odot}$ by the cosmological expansion. From our predictions we estimate an intrinsic, unlensed, chirp mass of $\simeq 10-12 M_\odot$, with a source redshift in the range $0.9<z<2.5$. We also outline a joint analysis over all baseline permutations that can stringently test our lensing interpretation of these two events. More generally, lensed events effectively multiply the number of baseline permutations and motivates the use of more interferometers for round the clock coverage of all repeat events of a given source, in order to maximise the orbital details and sky localization of lensed BBH sources.
\end{abstract}


\section{Introduction}
The all sky-coverage and current strong source sensitivity of LIGO/Virgo motivates a search for highly magnified, multiply-imaged GW sources that include two or more repeated events.  Strong gravitational lensing has been instrumental and invaluable for discovering and studying new populations of objects at high redshift, including the brightest known infrared galaxies detected in wide sky surveys \cite{Wardlow,Negrello,Bussmann} that are magnified by the general galaxy population with a median Einstein radius of $0.85\arcsec$.  A closely analogous situation should apply to the strongest LIGO/Virgo sources, which will be boosted by the same population of lensing galaxies, and can dominate over the unlensed events at the highest strains \cite{Broadhurst2018}. We have shown \cite{Broadhurst2018} that this is a significant effect for a BBH population following the mass distribution of known stellar mass black holes orbiting stars in our Galaxy, that is concentrated at $\simeq 10M_\odot$\cite{Remillard}, requiring a BBH event rate 2-3 orders of magnitude higher in range $1<z<4$ than today, and peaking at around $z\simeq 2$ like the measured cosmological rate of star formation\cite{Madau}.

The degeneracy between gravitational magnification ($\mu$) and luminosity distance \cite{Wang}
causes the luminosity distance to a GW source to be revised upwards by a factor of $\sqrt{\mu}$
and the inferred source frame masses of the compact objects to be revised down by a factor of $(1+z_o)/(1+z')$ where $z_o$ is the true redshift of the GW and $z'$ is the erroneously inferred redshift when lensing is ignored. Because many of the early GW detections appear to come from heavy black holes (\cite{Abbott et al. 2016a; Stevenson et al. 2017}), we have previously made an estimate of the expected distribution of chirp masses for galaxy lensing using the known optical depth of galaxy lenses with a magnification tail given by the universal form for caustics, $P(>\mu) \propto \mu^{-2}$ applied to the established log-normal distribution of stellar mass black holes in our Galaxy, and that is peaked around $10M_{\odot}$  as outlined in \cite{Broadhurst2018}. We have predicted an increase of $M_{Chirp}$ with magnification and redshift above $M_{Chirp}\geq 15M_{\odot}$ that match well the strong events detected (see Figures 2-3 of \cite{Broadhurst2018}). Furthermore, two LIGO events fall well below this relation with similar, chirp masses of $\simeq 10M_{\odot}$, as expected for unlensed local events. This event ratio of $\simeq 4/2$ in favor of lensed over unlensed events is reinforced by the addition of four more claimed events\cite{AbbotNEW} which all fall in our predicted lensing region of higher chirp mass, above $15M_{\odot}$, comprising a clear peak centered on $25M_{\odot}$, see Figure~3.  Amongst this magnified population, we have anticipated the discovery of repeated, multiply-lensed events for a significant minority $(\simeq 10-20\% $), given variation due to Earth rotating angular sensitivity of LIGO/Virgo \cite{Broadhurst2018}. At high magnification factors, lensing by fold caustics should dominate for small sources like BBH, producing two closely separated images and hence a relatively small time delay between repeated events dominated by elliptical galaxy lenses of days as shown in Figure~2.

\section{Comparison of Lensed Events}
 The case for strong lensing appears now by the detection of two apparently very similar waveforms for GW170814\cite{Abbot2} and the newly claimed GW170809\cite{AbbotNEW} which are separated by 5 days and 2 hours.  The waveforms agree in detail in Figure 1 (left and middle panels) including their phase, as shown by comparison with the models shown in Figure~3 (left and middle panels) and their sky positions are coincident within the uncertainties benefiting from the Virgo corroboration in the case of GW170814, with a relatively small estimated uncertainty of 87 sq degrees (90\% confidence area) and overlaps with the less well localized 340 sq degrees derived for GW170809 based on LIGO.
 
It is instructive to compare these waveforms with an unrelated event GW170818 of similar chirp mass (Figure~1) but well separated on the sky, that also benefits in positional accuracy from a Virgo detection \cite{Abbot2}, where an obvious phase difference is seen with respect to the other 2 events throughout the well detected last few orbits, of $\pi/2$ radians, as can be seen by comparison with the matching model waveform shown below in Figure~1 (generated with the open source software\cite{pythonCBC}).  

  The above frequency information provides the detector frame black hole chirp mass, $M_{\rm Chirp}(z)$ for both these events, that should be consistent with the same value for repeated images as the frequencies (and their derivatives, to first order) and phase depend only on the orbiting masses and cosmological redshift and not on "achromatic" lensing which is frequency independent. The LIGO/Virgo estimates of $27.1^{+1.6}_{-1.3}M_\odot$ and $30.0^{+2.5}_{-1.9}M_\odot$ agree well, confirming the visible similarity of the waveforms. These chirp mass estimates are made with independent best fitting templates, inevitably resulting in slightly different source parameters because of noise,
though most of the reported parameters are fully consistent at the 90\% confidence level \cite{AbbotNEW} for these two events. However, a joint fitting of parameters is now motivated to provide an explicit consistency test of the lensing hypothesis, starting from the raw data and applying established analysis methods under the prior assumption that they are identical as described below.

 The GW signal amplitudes of the two events can be compared to provide a relative magnification between the two events which should generally differ due to lensing. We can obtain this by using the published distance estimates for these two events because the luminosity distance is derived against predetermined waveform amplitudes predicted by General Relativity as a function of $M_{\rm Chirp}(z)$ ($m_1,m_2$). Since the measured strain depends inversely on luminosity distance $D$, and it scales as $\sqrt{\mu}$, then the relative magnification is $(\mu_1/\mu_2)=(D_2/D_1)^2$. Since the published distances are $990^{+320}_{-380}$ Mpc and $580^{+160}_{-210}$ Mpc, this leads to a magnification ratio of $2.8\pm0.9$, typical of other small lensed sources such as QSOs
and SNe. Note that these distances are translated by LIGO/Virgo into source redshifts of $z=0.2$ and $z=0.12$ respectively, for their respective luminosity distances, whereas a lensing based estimate is much higher $1<z<4$ for the distribution expected in our previous work\cite{Broadhurst2018}. 
The absolute level of magnification can also be estimated with reference to
our previous predictions, if one adopts as the most likely redshift a value consistent with the expected value from lensing the mean redshift is $z\simeq 1.2$ corresponding to a luminosity distance of 8000 Mpc. This can be compared with
the mean of the two estimated luminosity distances, $\approx 700\pm 150$ Mpc providing an approximate magnification $\mu\simeq 130$ 
that can be compared with empirically reliable estimates of magnifications for SNe sources of standard luminosity where a factor of up to $\simeq 50$ has been determined for individual lenses\cite{OguriSN,Goobar}. Furthermore, much higher magnifications of several thousand have also been derived for individually lensed high-z stars falling on the caustics of galaxy clusters \cite{Kelly,Diego,Dai1,Oguri}. 

\section{Processing the Dual Event Gravitational Lensed GW}
Here we consider what it would take to identify unambiguously a multiply-imaged black hole merger? When as a result of gravitational lensing there are two events for a single binary black hole merger, the data processing, modelling and fitting are more sophisticated as more information and baselines exist.
In the specific case here of GW170809 and GW170814 which are 5 days 2hours, 2 min and 21.7 minutes apart (5 days 2 .04 hours).
This means that the detectors are not at the same location for each event but are effectively rotated nearly 3406  cos$\lambda$  km  (2164 cos$\lambda$ miles) where  $\lambda$ is the latitude of the detector at LIGO Hanford (LHO) $46^\circ  27' 19"$ N, LIGO Livingston (LLO)	$30^\circ 33' 46"$ N, Virgo $43^\circ 37' 53"$ N.
Figure 4 shows an example of the configuration for the double event.



In this configuration we see that with multiple event lensing, LIGO/Virgo becomes a more true aperture synthesis interferometer.
Instead of one spatial baseline for LIGO or three baselines (though slightly degenerate) for LIGO/Virgo,
the configuration for these events has 6 possible (one nearly degenerate) spatial baselines and with Virgo adding another possible 4
or 8 if both events are observed by Virgo (and at lower signal to noise ratio a Virgo-Virgo baseline).
In other words, suddenly, up to 14 baselines could be available to have cross correlated wave forms and co-joined into a combined fit.
It is clear that a combined fit is possible and will give good effective chi-squared as the high-signal-to-noise portions of the two events 
waves forms match very well, though there is sufficient noise in the events, particularly GW170809 to allow a good joint fit.
The data processing can go forward straight forwardly treating the two events as independent measurements of the same event as though it comes
from the same physical location. 
We estimate that lensing is most likely a result of a large elliptical galaxy which is much more likely than from a galaxy cluster. In either case the image separation to be less than a couple of arcseconds as it must go through very near a caustic. 

This angular size is very much smaller than any angular resolution from the analysis of the events.
It is likely that a combined analysis of the data will show consistency with being a single black hole merger from the same place on the sky and provide additional improved constraints on the orbital parameters.
If the improvement in angular resolution is sufficient, then it would be possible to improve upon the current area that spans 5-10 degree within which multiple structures are identified in a deep DES related spectroscopic data including prominent over densities at z=0.06 and z=0.15\cite{LigoDES2019} and several clusters that can boost the optical depth to lensing and enhance the size of galaxy lensing caustics in this direction resulting a a more magnified Ligo/Virgo event. Currently even though GW170814 has a relatively small angular region 87 sq. degrees it still includes a large number of structures as shown in Figure 1 of \cite{LigoDES2019}. These structures could easily be the home galaxies of the BBH merger or the source of the high-magnification gravitational lensing.
However, as we claim, it is one of two gravitational lens images with GW170809, then the joint analysis should indicate a good fit and provide a more restricted sky coverage and thus allow one to pick out likely lensing elements.

\subsection{Microlensing in Gravitational Waves}
Concerns are sometimes expressed that microlensing might be an important effect in high magnification gravitational lensing of gravitational waves. For high magnification the two paths of the GW must come very near a critical curve.  
The corresponding caustic region might be disturbed by massive stars passing through and distorting the caustic providing both an issue whether 
diffraction is important and if there are any time dependent effects. 
However, the probability of a microlens disturbing the GW at LIGO frequencies is low. 
For effects to be significant at LIGO frequencies one would need stellar masses (stars or remnants) larger than 100 $M_\odot$ (Diego et al. 2019)
Such objects are rare and though it is possible that a GW intersects one of these massive objects near a critical curve, they are unlikely. The more common, but less massive microlenses with masses of a few or less solar masses can still produce interference effects but at frequencies higher than those probed by LIGO/Virgo. An exception is for GW that travel along the path with negative parity. As shown by Diego et al (2019), even moderate masses of order 1 solar mass can produce a modulation of the observed strain (and affect negatively their detectability, since these modulations are not included  in the data bank of filters used to detect the GW) when the wave is demagnified by the microlens. Microlenses on the side of the critical curve where the parity is negative, have a higher probability of demagnification than magnification so in these situations it would not be possible to detect a GW counter image event (Diego et al. 2019). 

\section{Conclusions}
We have pointed out the evident similarity of the LIGO/Virgo events GW170809 and GW170814 and how they are consistent with being the same binary black hole merger, under an appealing lensing interpretation. A fuller joint analysis is now feasible as described above with additional baselines providing an opportunity in this case to examine the consistency of lensing for this pair of events with up to 14 baselines that can sharpen the resulting waveforms by excluding more uncorrelated noise, leading to improved additional BBH orbital details. Note that we do not, in general, expect to catch many such pairs given the current interferometer configuration which has a strong and complex sky pattern that rotates daily with the Earth so that catching counter images will not become routine until new interferometers are operational, extending the longitudinal coverage.

Assuming our finding of about 2 to 1 lensed to unlensed events continues to hold up,
then it is most likely that the first observed NSBH merger will be lensed
and the lower mass object will appear to have about twice the mass of a typical neutron star due to redshift expansion of the waveform.
However, if the mass and spin of the BH are typical, then though there will be an optical signal,
it is likely significantly less than NSNS merger and might not be detected.\cite{Mukul}
It is therefore important to realize this is a possibility to alert  multi-messenger follow-up(particularly photons) that might be able to detect such a lensed NSBH event.

Irrespective of the validity of this viable case of multiple lensing, the four newly claimed 
LIGO/Virgo events all lie at high chirp masses reinforcing our earlier lensing interpretation based on the first 6 events, made on statistical grounds\cite{Broadhurst2018}, with no need for a new hypothetical high mass stellar progenitor population or primordial black hole explanation. This lensing solution implies an abundance of normal stellar mass black holes at $z\sim 2$, rather than $z\simeq 0.1$,  that we predict will continue to dominate GW detections for the foreseeable future until the sensitivities improves by an order of magnitude, to providing sufficient volume  for the unlensed  $\simeq 10 M_\odot$ peak of normal stellar mass black holes  to take over and dominate the detections at lower signal strength. This division of BBH events into local and cosmologically distant events is a bonus for gravitational wave astronomy as then the evolution of stellar BBH mass ratios, orbits and spins, can be traced over the Hubble time.



\subsection{ } T.J.B. thanks the IAS/HKUST for generous hospitality. The authors wish to thank  Alfred Amruth, Mandy Chen and Liang Dai for useful conversations. J. M. D. acknowledges the support of project AYA2015-64508-P(MINECO/FEDER, UE) and the hospitality of UPenn. G.F.S. thanks the IAS TT \& WF Chao Foundation for their support.

\subsection{}{All authors contributed significantly to this work.}

\subsection{Author Information} Correspondence and requests for materials should be addressed to tom.j.broadhurst@gmail.com, jdiego@ifca.unican.es, gfsmoot@lbl.gov

\subsection{Competing financial interests} The authors declare no competing financial interests.

\renewcommand{\figurename}{Figure}

\clearpage
\begin{figure}[ht]
\centering
\includegraphics[width=15.5cm]{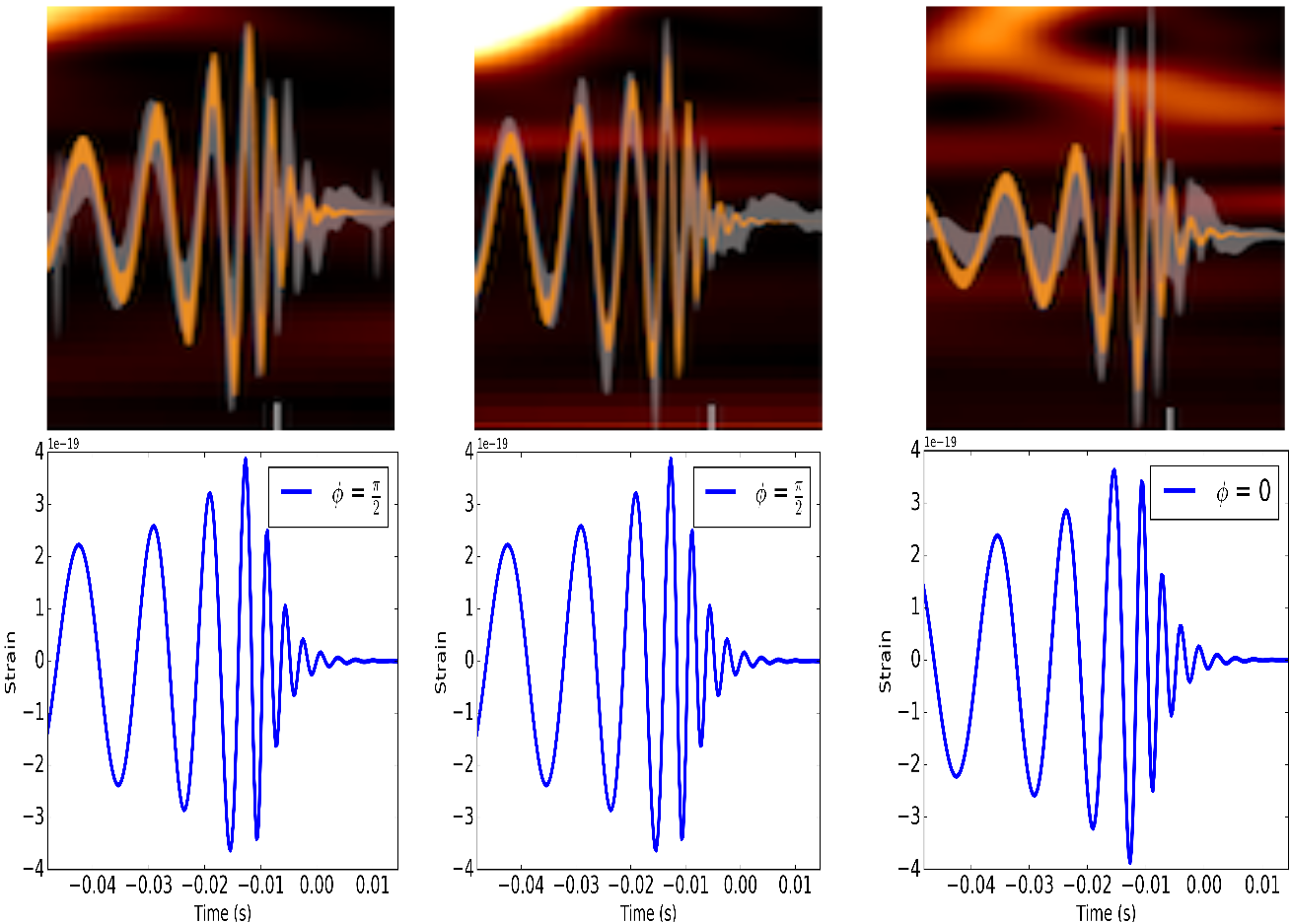}
\caption{{\bf Phase similarity:} The similar pair of detected waveforms of interest here are shown for GW170809 and GW170814 in the left and middle panels respectively\cite{AbbotNEW}. These are compared with model waveforms below showing that a fixed phase, defined with respect to the moment of merger shown underneath in blue, and fixed chirp mass of $28.5M_\odot$ (in the detector frame) consistent with the observed estimates generated by the pyCBC open software\cite{pythonCBC}, well matched to this pair of observed waveforms and demonstrating their similarity for a fixed detector chirp mass. For comparison, the righthand panel shows an independent event GW170818, that has a similar chirp mass ($31M_{\odot}$) but is well constrained in a different sky location with the benefit of a Virgo detection. This waveform is clearly different in appearance than the other two and readily explained by
a differing phase as can be seen from the model shown underneath where the phase is shifted by $\pi/2$ approximating well the form of this observed waveform.}
\label{fig:result1}
\end{figure}

\clearpage
\begin{figure}[ht]
\centering
\includegraphics[width=15.5cm]{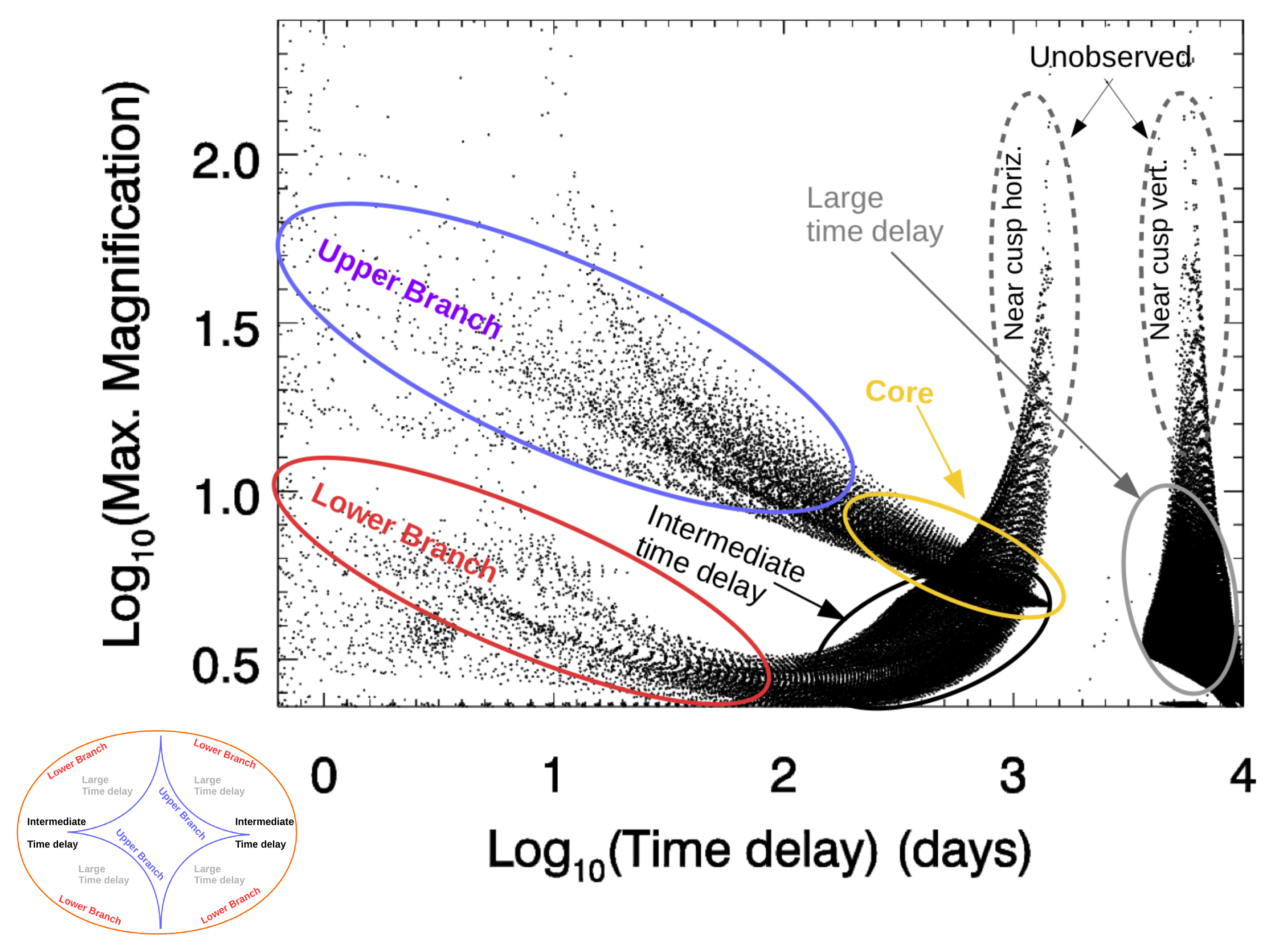}
\caption{{\bf Time Delay Between Pairs of Lensed Events:} This plot shows the predicted time delays between image pairs for a massive elliptical galaxy lens, at $z\approx 0.5$ and a source redshift of, $z\simeq2$. The y-axis shows the largest magnification of the pair, while the x-axis shows the time delay. The different {\it branches} correspond to different regions in the source plane where the tangential and radial caustics form a diamond shape and a surrounding  ellipsoid respectively. A cartoon of the source plane with the different regions is shown in the bottom corner. The region denoted "Upper branch" is the one that is relevant for this work (sort time delays and large magnification factors) and correspond to source positions that are close to the tangential caustic but on the interior of the diamond shape caustic. In this region, the typical time delay can be a few days and magnification can be a few tens to few hundreds. The region marked as "unobserved" correspond to those pairs of events for which one of the counter images has very small magnification  while the other has a higher magnification.
}
\label{fig:timedelay}
\end{figure}

\clearpage
\begin{figure}[ht]
\centering
\includegraphics[width=15.5cm]{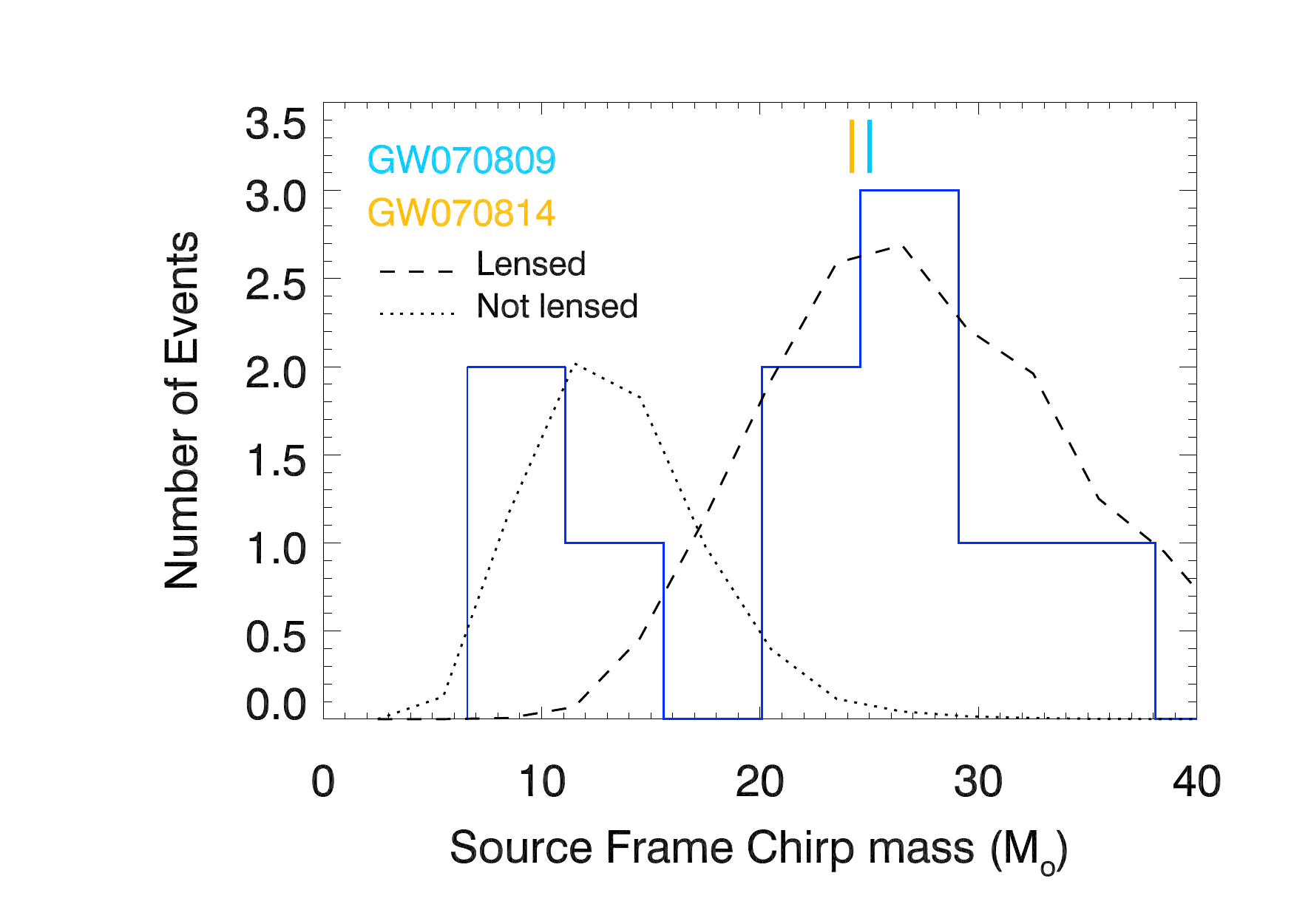}
\caption{{\bf Distribution of chirp masses and Lensing :} Here we
compare all currently claimed BBH chirp masses 
inferred by LIGO/Virgo\cite{AbbotNEW} with our previously  
lens model predictions\cite{Broadhurst2018} together with our normalized event rate predictions, comprising lensed events (dashed curve) 
and non-lensed events (dotted curve) that we have predicted or the current LIGO/Virgo sensitivity limit.
It is notable that the distribution of observed chirp masses has a distinctive peak just like our lensing predicted curve, corresponding 
to sources at $z>1$, with the exception of the 3 lowest mass events that match well the expected contribution from relatively nearby sources that are not lensed, with $z<0.3$. The new data suggests that the lensed events represent a majority. Assuming the  pair of events claimed are just one BBH merger doubly lensed this becomes a 2:1 ratio in favour of lensing.
} 
\label{fig:result3}
\end{figure}

\clearpage
\begin{figure}[ht]
\centering
\includegraphics[width=15.5cm]{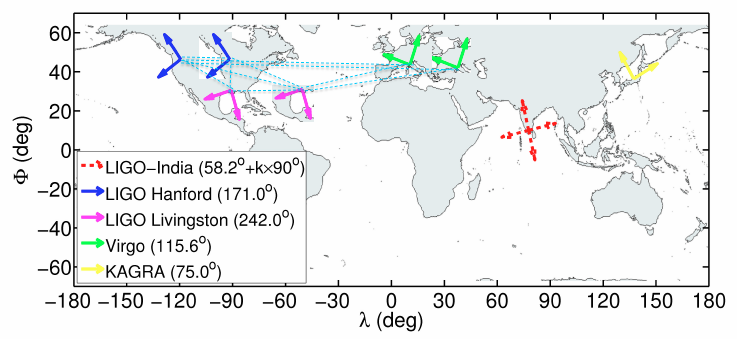}
\caption{{\bf Distribution of Effective Detectors:} This map shows how double event lensing produces an additional effective set of detectors which result in a large increase of additional baselines. The 2 events GW170809 and GW170814 are separated on the sky by about 30 degrees corresponding to the 5 days plus 2hrs difference in arrival time. This increase in detectors can then provide much more information on the BBH merger both in terms of orbital parameters and in localizing where it is on the sky. The four apparent LIGO detectors actually provide 6 spatial baselines and combined with observations from Virgo can result in as many as 14 total. A BBH merger with three images observed as events would provide three times the existing detectors and many times the baselines provided the arrival times had good spacings. 
}
\label{fig:result2}
\end{figure}

\end{document}